\documentclass[prd,reprint,showpacs,showkeys]{revtex4-1}
\usepackage{amsfonts}
\usepackage{mdframed}
\usepackage{amssymb}
\usepackage{amsmath}
\usepackage{graphicx}
\usepackage[font={footnotesize,it}]{caption}

\begin{document}

\preprint{}
\title{Cloud of strings as source in $2+1-$dimensional $f\left( R\right)
=R^{n}$ gravity}
\author{S. Habib Mazharimousavi}
\email{habib.mazhari@emu.edu.tr}
\author{M. Halilsoy}
\email{mustafa.halilsoy@emu.edu.tr}
\affiliation{Department of Physics, Eastern Mediterranean University, Gazima\u{g}usa,
Turkey.}

\begin{abstract}
We present three parameters exact solutions with possible black holes in $%
2+1-$dimensional $f\left( R\right) =R^{n}$ modified gravity coupled
minimally to a cloud of strings. These three parameters are $n,$ the cloud
of string coupling constant $\xi $ and an integration constant $C$. Although
in general one has to consider each set of parameters separately; for $n$
an even integer greater than one we give a unified picture providing black
holes. For $n\geq 1$ we analyze null / timelike geodesic within the context
of particle confinement.
\end{abstract}

\maketitle

\section{Introduction}

The advantages of working in lower-dimensional gravity, specifically in $%
2+1- $dimensions has been highlighted extensively during the recent decades.
The interest started all with the discovery of $2+1-$dimensional black hole
solution by Banados, Teitelboin and Zanelli (BTZ) \cite{BTZ}. The physical
source of the BTZ black hole was a cosmological constant which was
subsequently extended with the presence of different sources \cite{BTZ2}.
Physically how significant are these sources?. Addition of scalar \cite%
{Scalar} and electromagnetic, both linear and non-linear fields has almost
been a routine while exotic and phantom fields also found room of
applications in the problem. A source that is less familiar is a cloud of
strings \cite{SC} which was considered in Einstein's general relativity.
Within this context in $3+1-$dimensions the importance of a string cloud has
been attributed to the action-at-a distance interaction between particles.
For a detailed geometrical description of a string cloud we refer to \cite%
{SC}. In this study we extend such a source to the $f\left( R\right) =R^{n}$
gravity which is a modified version of general relativity \cite{Rn}. In $D-$%
dimensional spacetime the energy-momentum tensor for a string cloud is
represented by $T_{\mu }^{\nu }=\frac{\xi }{r^{D-2}}diag\left(
1,1,0,0,...,0\right) $ where $\xi $ is a positive constant. In $3-$%
dimensions, which will be our concern here this amounts to $%
T_{0}^{0}=T_{1}^{1}=\frac{\xi }{r},$ with $T_{2}^{2}=0$ where our labeling
of coordinates is $x^{\mu }=\left( t,r,\theta \right) .$ Compared with the
energy-momentum of the scalar and electromagnetic sources which are of the
order $\frac{1}{r^{2}}$ in $2+1-$dimensions the order $\frac{1}{r}$ for a
string cloud may play an important role. Briefly the singularity at $r=0$ is
weaker in comparison with different sources. This makes the motivation for
us to conduct the present study. We note that our cloud of strings is
reminiscent of the wormhole "fur coat" model in $5D$ Reissner-Nordstr\"{o}m
black hole \cite{Dzh}.

Such a cloud of $1-$dimensional string play the role of particles in analogy
with the $1-$dimensional gas atoms. The spatial geometry is confined to the
polar plane with the cyclic angular coordinate. With the addition of time
variable the sheet description of the string becomes more evident. The
strings are open, originating at the singular origin and extending to
infinity, vanishing with $r\rightarrow \infty .$ Thus, for $r\rightarrow
\infty $ our model reduces to the source-free (vacuum) $f\left( R\right) $
model, which derives its power from the curvature of geometry. For $\xi =0$
in $f\left( R\right) =R$ model the spacetime is automatically flat unless
supplemented by other sources. In $f\left( R\right) $ gravity, on the other
hand even though we can take $\xi =0$ we have still room for a non-flat
metric albeit this may not be a black hole.

We investigate the field equations of $f\left( R\right) $ gravity in the
presence of a string cloud. In general, these are highly non-linear
differential equations but owing to the simplicity of our source and the $%
2+1-$dimensions we are fortunate to obtain a large class of exact solutions.
For particular choice of the parameter $n$ our solution can be interpreted
as black holes, while for other choices it corresponds to cosmological
models. 

\section{String cloud source in $f\left( R\right) =R^{n}$ gravity}

Let us start with the action of the $2+1-$dimensional $f\left( R\right) -$%
gravity coupled to the cloud of strings%
\begin{equation}
I=\frac{1}{16\pi G}\int d^{3}x\sqrt{-g}f\left( R\right) +I_{S}
\end{equation}%
in which $f\left( R\right) =R^{n}$ is a function of Ricci scalar $R,$ $n$ is
a real constant and 
\begin{equation}
I_{S}=\int_{\Sigma }m\sqrt{h}d\lambda ^{0}d\lambda ^{1}
\end{equation}%
where $\lambda ^{A}$, $(A=0,1)$ are the string parameters. The world-sheet
bivector is defined by 
\begin{equation}
\sum\nolimits^{\mu \nu }=\epsilon ^{AB}\frac{\partial x^{\mu }}{\partial
h^{A}}\frac{\partial x^{\nu }}{\partial h^{B}}
\end{equation}%
in which $\epsilon ^{01}=1=-\epsilon ^{10}$ is the $2-$dimensional
Levi-Civita symbol. Note that the string metric has the line element%
\begin{equation}
ds_{s}^{2}=h_{AB}dh^{A}dh^{B}.
\end{equation}%
Accordingly the corresponding energy-momentum tensor of the string-cloud is 
\begin{equation}
T^{\mu \nu }=\rho \frac{\sum\nolimits^{\mu \alpha }\sum\nolimits_{\alpha
}^{\nu }}{\sqrt{-h}}
\end{equation}%
where $\rho $ is the energy-density and $h=\det \left( h_{AB}\right) $
refers to the world sheet of the string. The latter expression for $T^{\mu
\nu }$ is meaningful as long as $h<0.$

Variation of the action $I$ with respect to $g_{\mu \nu }$ provides the
field equations (in metric formalism)%
\begin{equation}
\frac{df}{dR}R_{\mu }^{\nu }-\frac{1}{2}f\delta _{\mu }^{\nu }-\nabla ^{\nu
}\nabla _{\mu }\frac{df}{dR}+\delta _{\mu }^{\nu }\square \frac{df}{dR}%
=T_{\mu }^{\nu }
\end{equation}%
in which $\square $ is the covariant Laplacian and 
\begin{equation}
T_{\mu }^{\nu }=\frac{\xi }{r}diag.\left[ 1,1,0\right]
\end{equation}%
is the energy momentum tensor of the cloud of string in which $\xi $ is a
real constant. The static and circular symmetric line element is set to be%
\begin{equation}
ds^{2}=-U\left( r\right) dt^{2}+\frac{1}{V\left( r\right) }%
dr^{2}+r^{2}d\theta ^{2}
\end{equation}%
with two unknown functions $U\left( r\right) $ and $V\left( r\right) .$ The
field equations explicitly written are 
\begin{multline}
\frac{df}{dR}R_{t}^{t}-\frac{f}{2}+V\left( R^{\prime 2}\frac{d^{3}f}{dR^{3}}+%
\frac{d^{2}f}{dR^{2}}R^{\prime \prime }\right) + \\
\left( \frac{V^{\prime }}{2}+\frac{V}{r}\right) R^{\prime }\frac{d^{2}f}{%
dR^{2}}=\frac{\xi }{r},
\end{multline}%
\begin{equation}
\frac{df}{dR}R_{r}^{r}-\frac{f}{2}+\left( \frac{VU^{\prime }}{2U}+\frac{V}{r}%
\right) R^{\prime }\frac{d^{2}f}{dR^{2}}=\frac{\xi }{r}
\end{equation}%
and%
\begin{multline}
\frac{df}{dR}R_{\theta }^{\theta }-\frac{f}{2}+V\left( R^{\prime 2}\frac{%
d^{3}f}{dR^{3}}+\frac{d^{2}f}{dR^{2}}R^{\prime \prime }\right) + \\
\left( \frac{VU^{\prime }}{2U}+\frac{V^{\prime }}{2}\right) R^{\prime }\frac{%
d^{2}f}{dR^{2}}=0
\end{multline}%
in which%
\begin{equation}
R_{t}^{t}=-\frac{\left[ V^{\prime }U^{\prime }Ur+2VU^{\prime \prime
}Ur-VU^{\prime 2}r+2VU^{\prime }U\right] }{4rU^{2}},
\end{equation}%
\begin{equation}
R_{r}^{r}=-\frac{\left[ V^{\prime }U^{\prime }Ur+2VU^{\prime \prime
}Ur-VU^{\prime 2}r+2V^{\prime }U^{2}\right] }{4rU^{2}}
\end{equation}%
and%
\begin{equation}
R_{\theta }^{\theta }=-\frac{\left( rVU^{\prime }+rUV^{\prime }\right) }{%
2r^{2}U}.
\end{equation}%
Note that, a prime stands for the derivative with respect to $r$. A set of
solutions to the above field equations are given by%
\begin{equation}
U\left( r\right) =r^{2\left( n-1\right) }\left( C-\frac{2n^{2}\xi r^{\frac{%
1-2n+2n^{2}}{n}}}{\alpha \left( 1-2n+2n^{2}\right) \left( 4n^{2}-6n+1\right) 
}\right)
\end{equation}%
and%
\begin{equation}
V\left( r\right) =r^{\frac{2\left( 1-n\right) \left( 2n-1\right) }{n}%
}U\left( r\right) ,
\end{equation}%
in which $C$ is an integration constant and%
\begin{equation}
\alpha ^{n}=n^{n}\left( \frac{4n^{2}-6n+1}{4\left( 2n-1\right) \xi }\right)
^{1-n}.
\end{equation}%
To complete the set of solutions we provide also the explicit form of the
Ricci scalar $R$ which reads as%
\begin{equation}
R=\frac{4\xi n\left( 2n-1\right) }{\left( 4n^{2}-6n+1\right) \alpha r^{\frac{%
1}{n}}}.
\end{equation}%
Let's note that $n\neq 1/2$ is a real constant which must exclude also the
values $\frac{3\pm \sqrt{5}}{4}$ since they are the roots of the denominator
of $R$.

We would like to add also that, due to an arbitrary value for $n$ and $\xi $
our solution is a three parameter solution. For any specific value of $n$
one has to choose an appropriate sign for $\xi $ which in cases both sign
may be acceptable. Setting $n$ and the sign of $\xi $ give us an equation
for $\alpha $ given by (17). Depending on the number of real solutions this
equation may admit, we will get different metric functions. For instance
let's consider $n=2.$ In this case one finds $\alpha ^{2}=4\left( \frac{%
12\xi }{5}\right) $ which imposes $\xi >0$ and consequently there are two
solutions for $\alpha $ given by $\alpha =\pm 2\sqrt{\frac{12\xi }{5}}.$ For
positive/negative $\alpha $ one finds%
\begin{equation}
U_{\pm }\left( r\right) =r^{2}\left( C\mp \frac{2\sqrt{15\xi }r^{\frac{5}{2}}%
}{75}\right) .
\end{equation}%
Here clearly $U_{-}\left( r\right) $ is a black hole solution considering $%
C<0$ and therefore we may rewrite the metric as%
\begin{equation}
U_{-}\left( r\right) =r^{2}\left\vert C\right\vert \left( \left( \frac{r}{%
r_{h}}\right) ^{\frac{5}{2}}-1\right)
\end{equation}%
in which the horizon is located at%
\begin{equation}
r_{h}=\left( \frac{75\left\vert C\right\vert }{2\sqrt{15\xi }}\right) ^{%
\frac{2}{5}}.
\end{equation}%
In general, for $n\geq 2$ an even integer number the picture is the same as $%
n=2$ i.e., $\alpha =\pm n\left( \frac{4n^{2}-6n+1}{4\left( 2n-1\right) \xi }%
\right) ^{\frac{1-n}{n}}$ with $\xi >0.$ The general solution hence can be
cast as a black hole solution provided $C<0$ and negative $\alpha $ such that%
\begin{equation}
U_{-}\left( r\right) =r^{2\left( n-1\right) }\left\vert C\right\vert \left(
\left( \frac{r}{r_{h}}\right) ^{\frac{1-2n+2n^{2}}{n}}-1\right)
\end{equation}%
in which the location of the horizon is given by%
\begin{equation}
r_{h}=\left( \frac{2\left( 1-2n+2n^{2}\right) \left( 2n-1\right) \left\vert
C\right\vert }{n\left( \frac{4\left( 2n-1\right) }{4n^{2}-6n+1}\xi \right) ^{%
\frac{1}{n}}}\right) ^{\frac{n}{1-2n+2n^{2}}}.
\end{equation}%
For odd integers and other real numbers as we mentioned above, one must
carefully go through any individual case and find the final form of the
spacetime. Ultimately it gives either a black hole solution as we mentioned
above or a particle solution. For instance, in the case of $n=2$ the
positive branch i.e., $U_{+}\left( r\right) $ with positive $C$ the solution
represents a particle solution \cite{Dzh2}.

\subsection{$f\left( R\right) =R$ with geodesics}

In the second example we consider $n=1$ which simply gives the Einstein's $R$%
-gravity coupled to the cloud of string minimally. Let's note that this
solution was found first by Bose, Dadhich and Kar in \cite{BDK}. The
solution becomes%
\begin{equation}
U\left( r\right) =V\left( r\right) =-M+2\xi r
\end{equation}%
and%
\begin{equation}
R=-\frac{4\xi }{r}.
\end{equation}%
As one observes, there is no restriction on the sign of $\xi $ and therefore
the solution represents either a singular black hole or a naked singularity.
In the case of the black hole the singularity is hidden behind an event
horizon located at%
\begin{equation}
r_{h}=\frac{M}{2\xi }.
\end{equation}

\subsubsection{Geodesics confinement for $f\left( R\right) =R$}

The black hole solution in $R$-gravity given by (24) is rather interesting
if we assume $M,\xi >0$. In this section we investigate the null and
timelike geodesics of this solution. Let's start with the Lagrangian 
\begin{equation}
L=\frac{1}{2}g_{\mu \nu }\frac{dx^{\mu }}{d\lambda }\frac{dx^{\nu }}{%
d\lambda }
\end{equation}%
in which $\lambda $ is an arbitrary parameter. We also introduce the null ($%
\epsilon =0$) and timelike ($\epsilon =1$) geodesics by%
\begin{equation}
g_{\mu \nu }\frac{dx^{\mu }}{d\lambda }\frac{dx^{\nu }}{d\lambda }=-\epsilon
.
\end{equation}%
The line element%
\begin{equation}
ds^{2}=-Udt^{2}+\frac{dr^{2}}{U}+r^{2}d\theta ^{2}
\end{equation}%
in which $U=-M+2\xi r$ admits two Killing vectors, i.e., $\partial _{t}$ and 
$\partial _{\theta }$ corresponding to the conserved energy and angular
momentum given by%
\begin{equation}
E=-U\frac{dt}{d\lambda }
\end{equation}%
and%
\begin{equation}
\ell =r^{2}\frac{d\theta }{d\lambda }.
\end{equation}%
Considering these one finds the only equation to be solved as%
\begin{equation}
\left( \frac{dr}{d\lambda }\right) ^{2}+U\left( \epsilon +\frac{\ell ^{2}}{%
r^{2}}\right) =E^{2}.
\end{equation}%
For the null radial geodesics with $\epsilon =\ell ^{2}=0$ one finds%
\begin{equation}
r_{\pm }=r_{0}e^{\pm 2\xi \left( t-t_{0}\right) }+r_{h}\left( 1-e^{\pm 2\xi
\left( t-t_{0}\right) }\right)
\end{equation}%
in which $r_{0}$ and $t_{0}$ are the initial location and time of the null
particle with respect to a distant observer. The sign $\pm $ stands for the
direction of the initial velocity with positive for outward and negative for
inward motion. As one can see 
\begin{equation}
\frac{dr_{\pm }}{dt}=\pm 2\xi \left( r_{\pm }-r_{h}\right)
\end{equation}%
and 
\begin{eqnarray}
\lim_{t\rightarrow \infty }r_{+} &=&\infty \\
\lim_{t\rightarrow \infty }\frac{dr_{+}}{dt} &=&\infty  \notag
\end{eqnarray}%
while%
\begin{eqnarray}
\lim_{t\rightarrow \infty }r_{-} &=&r_{h}. \\
\lim_{t\rightarrow \infty }\frac{dr_{-}}{dt} &=&0.  \notag
\end{eqnarray}%
In the case of timelike radial geodesics with $\epsilon =1,\ell ^{2}=0$ one
finds (for $\lambda =\tau $)%
\begin{equation}
r=r_{0}-\frac{1}{2}\xi \tau ^{2}
\end{equation}%
in which $\tau $ is the proper time and $r_{0}$ is the initial position of
the particle. To reach the horizon, the proper time needed by the particle
is just $\tau _{h}=\sqrt{\frac{2\left( r_{0}-r_{h}\right) }{\xi }}$In terms
of the coordinate time $t$ the equation of motion becomes%
\begin{equation}
\left( \frac{dr}{dt}\right) ^{2}=U^{2}\left( 1-\frac{U}{E^{2}}\right) .
\end{equation}%
If we consider the particle is at rest at $t=t_{0}$ we find $E^{2}=M\left( 
\frac{r_{0}}{r_{h}}-1\right) $ that consequently implies%
\begin{equation}
\left( \frac{dr}{dt}\right) ^{2}=4\xi ^{2}\left( r-r_{h}\right) ^{2}\left( 
\frac{r_{0}-r}{r_{0}-r_{h}}\right)
\end{equation}%
which admits the solution%
\begin{equation}
r=r_{0}-\left( r_{0}-r_{h}\right) \tanh ^{2}\left[ \xi \left( t-t_{0}\right) %
\right] .
\end{equation}%
Clearly the particle is attracted toward the black hole and when $%
t\rightarrow \infty $ the limit goes to $r=r_{h}.$ This time is not
comparable with the proper time interval needed for the particle to cross
the horizon.

In circular motion for a null particle one finds the only unstable orbit is
located at $r_{c}=2r_{h}$ (photon circle's radius) at which for the null
particle one finds 
\begin{equation}
\frac{E^{2}}{\ell ^{2}}=\frac{\xi }{2r_{h}}.
\end{equation}%
This can easily be justified from Eq. (32), by considering a potential of
the particle given by 
\begin{equation}
V\left( r\right) =U\left( r\right) \left( \epsilon +\frac{\ell ^{2}}{r^{2}}%
\right) -E^{2}
\end{equation}%
whose derivative satisfies $\frac{dV}{dr}=0$, at $r=r_{c},$ and $\epsilon =0$
for the null geodesics. For the timelike geodesics the similar analysis
applies with the substitution $\epsilon =1.$

Furthermore, for a timelike particle we find the stable orbit at $r=r_{c}$
where the angular momentum and the energy of the particle are given by%
\begin{equation}
\ell ^{2}=\frac{r_{c}^{3}}{r_{c}-2r_{h}}
\end{equation}%
and%
\begin{equation}
E^{2}=4\xi \frac{\left( r_{c}-r_{h}\right) r_{h}}{\left( r_{c}-2r_{h}\right) 
}.
\end{equation}%
We see clearly that $r_{c}>2r_{h}$ which is outside a photon circle around
the black hole.

\subsubsection{Generalization to $f\left( R\right) =R^{n}$ with $n$ an even
integer and $n\geq 2$}

The geodesic analysis given in the previous part can be generalized with $%
f\left( R\right) =R^{n}.$ As we are interested in the black hole geodesics,
we impose $n$ to be even integer bigger than one i.e., $n\geq 2.$ In order
to keep the mathematical expression analytic we only consider the radial,
null-geodesics. Accordingly one finds from the Euler-Lagrange equations%
\begin{equation}
\pm \frac{dr}{d\lambda }=Er^{\frac{\left( 1-n\right) \left( 2n-1\right) }{n}}
\end{equation}%
and%
\begin{equation}
\pm \frac{dt}{d\lambda }=\frac{E}{U}
\end{equation}%
in which $E$ is the energy of the particle and $\lambda $ is an affine
parameter. Combining these two equations we find%
\begin{equation}
\frac{dr}{dt}=\pm r^{\frac{\left( 1-n\right) \left( 2n-1\right) }{n}}U
\end{equation}%
with its integral expression in the form%
\begin{equation}
\int_{r_{0}}^{r}\frac{r^{\frac{\left( n-1\right) \left( 2n-1\right) }{n}}dr}{%
U}=\pm \left( t-t_{0}\right)
\end{equation}%
where $t_{0}$ is the initial time when the particle is located at $%
r_{0}>r_{h}$ and $r_{h}$ is the event horizon. Plugging (22) into the latter
equation we reexpress%
\begin{equation}
\int_{r_{0}}^{r}\frac{r^{\frac{1-n}{n}}dr}{\left\vert C\right\vert \left(
\left( \frac{r}{r_{h}}\right) ^{\frac{1-2n+2n^{2}}{n}}-1\right) }=\pm \left(
t-t_{0}\right) .
\end{equation}%
Obviously an exact integral of this expression is out of our reach. Even for
particular values of $n$ we have to appeal to the asymptotic behaviors.
Considering $r$ to be large i.e., $\frac{r_{h}}{r}\ll 1,$ and $n>0$ one
approximately finds%
\begin{equation}
r^{2\left( 1-n\right) }-r_{0}^{2\left( 1-n\right) }\simeq \mp \omega \left(
t-t_{0}\right)
\end{equation}%
in which the constant $\omega $ is given by%
\begin{equation}
\omega =2\left( n-1\right) \left\vert C\right\vert \left( \frac{1}{r_{h}}%
\right) ^{\frac{1-2n+2n^{2}}{n}}.
\end{equation}%
Let's add that considering $\frac{r_{h}}{r}\ll 1$ in (49) implicitly implies
that the initial location of the particle is far from the horizon i.e., $%
\frac{r_{h}}{r_{0}}\ll 1$ too. Latter equation implies that at least for
this specific choices the motion is confined.

\section{Conclusion}

In search of alternative black holes to the BTZ in $2+1-$dimensions a
particular case was considered in Einstein's theory of $f\left( R\right) =R$
in which the source is a cloud of string \cite{BDK}. Projected in the polar
plane the string can be considered as radial lines originating at the origin
and extending in radial in/out direction. This excludes the possibility of
closed string in such a geometry. The advantage of such a choice of source
becomes evident when substituted into the complicated $f\left( R\right)
=R^{n}$ gravity which we consider here. In other words our geometry is
powered by such a cloud of string in the $f\left( R\right) =R^{n}$ gravity
with the energy-momentum tensor $T_{0}^{0}=T_{1}^{1}=\frac{\xi }{r}$ and $%
T_{2}^{2}=0$ with $\xi =$constant. Obviously this satisfies the null energy
conditions. Although $n$ can be arbitrary in principle there are
restrictions on the choice of $n$ for a meaningful expression. For instance, 
$n=\frac{1}{2}$ and $n=\frac{3\pm \sqrt{5}}{4}$ are to be excluded in the
class of metric solutions. Depending on the other values of $n$ we obtain an
infinite class of possible metrics that describe black holes / naked
singularities in $2+1-$dimensional $f\left( R\right) =R^{n}$ gravity theory.
An interesting physical conclusion to be drawn from these solutions is the
role that the power $n$ plays in the confinement of (especially) null
geodesics. Although the most general analytic integration of the geodesics
is lacking we obtain an approximate connection between the parameter $n$ and
the confinement of the null geodesics for $n$ an even integer greater than
one. Whether a similar relation occurs in higher dimensions $(D>3)$ for a
cloud of strings as source remains to be seen.

\bigskip

\end{document}